\documentclass[showpacs,prb,superscriptaddress,twocolumn]{revtex4-1}
\usepackage{amsmath,amssymb}
\usepackage{graphicx}
\usepackage{multirow}
\usepackage{cancel}
\usepackage{dcolumn}
\usepackage{textcomp}
\usepackage{color} 

\usepackage[colorlinks,plainpages=false,linkcolor=blue,urlcolor=blue,citecolor=blue,pdfpagemode=UseNone,pdfstartview=FitBH]{hyperref}

\begin{document}

\title{A combined experimental and computational study of the pressure dependence of the vibrational spectrum of solid picene C$_{22}$H$_{14}$}

\author{F. Capitani} 
\affiliation{Dipartimento di Fisica, Universit\`{a} ~di Roma Sapienza, 
P.le Aldo Moro 2, 00185 Roma, Italy }

\author{M. H\"oppner}
\affiliation{Max Planck Institute for Solid State Research, Heisenbergstrasse 1, D-70569, Stuttgart, Germany}

\author{B. Joseph} 
\affiliation{Dipartimento di Fisica, Universit\`{a} ~di Roma Sapienza, 
P.le Aldo Moro 2, 00185 Roma, Italy }

\author{L. Malavasi} 
\affiliation{Dipartimento di Chimica, Universit\`{a}  di Pavia, Via Taramelli 16, 27100 Pavia, Italy}

\author{G.A. Artioli} 
\affiliation{Dipartimento di Chimica, Universit\`{a}  di Pavia, Via Taramelli 16, 27100 Pavia, Italy}

\author{L.~Baldassarre}\affiliation{Center for Life NanoScience@Sapienza, Istituto Italiano di Tecnologia, Viale Regina Elena 291, Roma, Italy} 

\author{A. Perucchi}
\affiliation{Sincrotrone Trieste, S.C.p.A., Area Science Park, I-34012, Basovizza, Trieste, Italy}

\author{M. Piccinini}\altaffiliation{Present address: ENEA, C.R. Frascati, Via E. Fermi, 45, 00044 Frascati (Rome), Italy} 
\affiliation{Porto Conte Ricerche S.r.l., SP 55 km 8.400 Loc. Tramariglio, 07041 Alghero (SS), Italy}

\author{S. Lupi}
\affiliation{CNR-IOM and Dipartimento di Fisica, Universit\`{a} di Roma Sapienza, P.le Aldo Moro 2, 00185 Roma, Italy}

\author{P. Dore}
\affiliation{CNR-SPIN and Dipartimento di Fisica, Universit\`{a} ~di Roma Sapienza, P.le Aldo Moro 2, 00185 Roma, Italy }

\author{L. Boeri}
\affiliation{Max Planck Institute for Solid State Research, Heisenbergstrasse 1, D-70569, Stuttgart, Germany}
\affiliation{Institute of Theoretical and Computational Physics, TU Graz, Petersgasse 16, 8010, Graz, Austria}

\author{P. Postorino}\email[corresponding author: ]{Paolo.Postorino@roma1.infn.it} 
\affiliation{CNR-IOM and Dipartimento di Fisica, Universit\`{a} di Roma Sapienza, P.le Aldo Moro 2, 00185 Roma, Italy}

\date{\today}

\newcommand{\changes}[1]{{\color{red} #1}}

\begin{abstract}
We present high-quality optical data and density functional perturbation theory calculations for the vibrational spectrum of solid picene (C$_{22}$H$_{14}$) under pressure up to 8~GPa. 
First-principles calculations reproduce with a remarkable accuracy the pressure effects on both frequency and intensities of the observed phonon peaks.
We use the projection on molecular eigenmodes
to unambiguously fit the experimental spectra, resolving complicated
spectral structures, in a system with hundreds of  phonon modes.
With these projections, we can also quantify the loss
of molecular character under pressure. 
Our results indicate that picene, despite a $\sim 20 \%$ compression of the unit~cell, 
remains substantially a molecular solid up to 8~GPa, with phonon modes displaying a smooth and
uniform hardening with pressure, without any evidence of structural phase transitions.
The Gr\"uneisen parameter of the 1380~cm$^{-1}$ $a_1$ Raman peak
($\gamma_p=0.1$) is much lower than the effective value ($\gamma_d=0.8$)
due to K doping. Therefore, doping and pressure have very
different effects and it can be argued that softening of the 1380~cm$^{-1}$
mode is probably due to coupling with electronic states in K doped solid picene.
 \end{abstract}


\maketitle

\section*{Introduction}

\begin{figure}[h!btp]
\includegraphics[width=7.5cm]{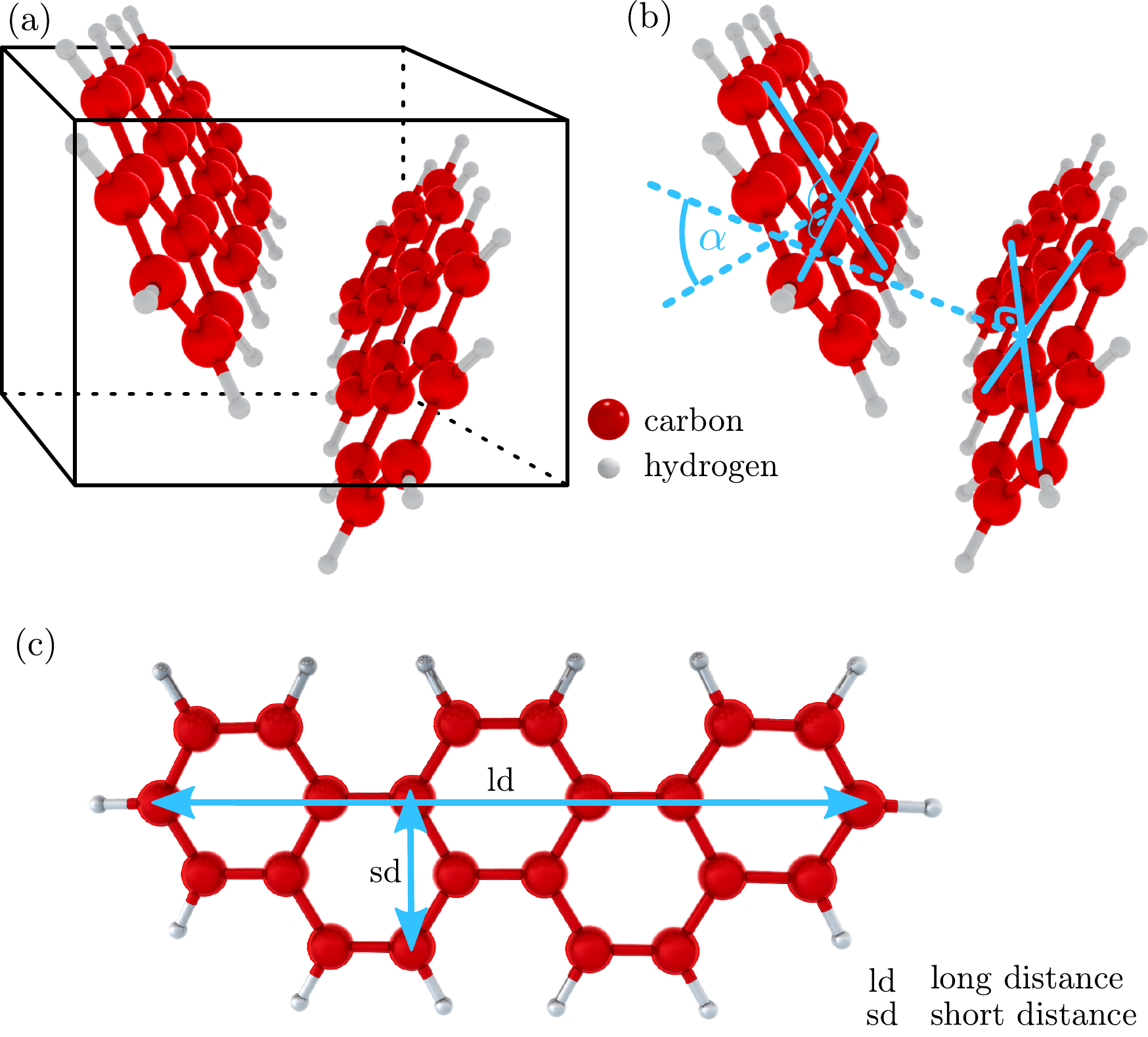}
   \caption{(Color online) Crystal structure of solid picene. (a)~Unit cell, comprising two molecules. (b)~Definition of distance and angle $\alpha$
   between the molecules. (c)~Top view of a single picene molecule (C$_{22}$H$_{14}$) and definition of the intramolecular distances. Carbon and hydrogen atoms are shown in red and white, respectively.}
\label{figs:1}
\end{figure}

A superconductive phase has recently been observed in potassium-doped picene,
 below a critical temperature (T$_c$) of 7-18 K.
\cite{Mitsuhash2010} This finding has attracted wide attention being the first report of ``high-$T_c$" superconductivity (SC) in an aromatic compound. Picene (C$_{22}$H$_{14}$) is indeed an alternant polycyclic aromatic 
hydrocarbon (PAH), {\em i.e.} a planar aromatic molecule, formed by juxtaposed benzene rings (see Fig. \ref{figs:1}). Specifically, this molecule comprises five rings, arranged in a zig-zag fashion. After the initial report, superconductive phases were also observed in other 
PAHs upon doping with alkali, alkali-earths and rare-earths. 
These other compounds, {\em i.e.} phenanthrene, \cite {Wang2011_SC-phenanthrene,Wang2011_SC-phenanthrene2, Wang2012_SC-phenanthrene3} coronene, \cite{Mitamura2011_perpestive-picene} and dibenzopentacene \cite{Xue2011_SC-dipicene} comprise three, six and seven benzene rings respectively, thus suggesting that
PAHs most likely form a completely new and possibly large class of superconductors.

The superconducting mechanism in PAHs is still a matter of theoretical and experimental debate;
first-principles studies have concentrated on picene, and shown that both the electron-electron ($ee$)~\cite{Giovannetti2011_electronicStr,Kim2011_electronicStr,valenti_2012,arita_2012,kim_2013}
and electron-phonon ($ep$) interaction~\cite{Casula2011_ele-ph,Subedi2011_ele-ph,Kato2011_ele-ph,Sato2012_ele-ph} 
are sizeable in these $\pi$-bonded systems.
Works that ascribe superconductivity to the $ep$ interaction disagree on the relative
importance of intermolecular, intramolecular and intercalant phonons to the $ep$ coupling.
In fact, linear-response calculations in the rigid-band approximation for solid picene find
that the coupling is dominated by intramolecular phonons, with an intensity 
($V_{ep}=\lambda/N(0)= 150$ meV for holes, and 110 meV for electrons),~\cite{Subedi2011_ele-ph} 
that is consistent with the results of simpler 
vibrational analysis.~\cite{Kato2011_ele-ph,Sato2012_ele-ph}
Linear response calculations which include the dopants explicitly in a theoretically-optimized
structure find a rather different spectral distribution of the  $ep$
coupling, which indicates an important role of the dopant and metallic screening. 
 This would make PAHs substantially different from the fullerenes,
where the existing models for superconductivity rely on local (molecular) estimates of the $ep$ coupling.~\cite{Massimo, Olle} 
A definite confirmation of this scenario could only come from a detailed comparison of measured and calculated phonon spectra for doped and undoped samples, which would allow to test the reliability of the approximations used in different first-principles calculations.

At the moment there are still difficulties in obtaining single phase, well characterized 
superconducting PAHs, while, due to recent advances in chemical synthesis,
high-quality samples of pure solid picene are now available.
Besides superconductivity, picene, like other PAHs, has important possible applications in organic electronics,~\cite{Mitamura2011_perpestive-picene,Okamoto2008_FET-picene}
and this has stimulated several experimental and theoretical studies of its  vibrational,~\cite{JPCM2012_IR_Ram_picene,Girlando2012_IR_Ram_picene,Kambe_2013} optical~\cite{Citroni2013,Fanetti2013_HP,Knupfer2011} and electronic~\cite{Xin2012_ARPES} properties.

In this work, we present a combined experimental and theoretical analysis of the vibrational
spectra of solid picene under pressure, up to 8~GPa.
We collected high-quality Raman and infrared (IR) data at room temperature, and compared
those with detailed linear-response calculations of the same spectra~\cite{Baroni2001_DFPT}
 as a function of pressure. Experimental and theoretical spectra showed a remarkable agreement, for both the
positions and intensities of the peaks. We also propose a novel theoretical analysis of the vibrational modes, based on the projections
on molecular eigenvectors, which permits us to obtain a deep microscopical insight into
the phonon spectra. We show that, besides permitting to disentangle the complicated experimental
spectra, this analysis can also be used to characterise
the increase of intermolecular interactions as a function of pressure.
A similar analysis, applied to other pure and intercalated 
molecular solids, could help to quantify intermolecular interactions.

The main outcomes of our study are: (a)~a~complete characterization of the phonon spectrum of picene, (b)~a~detailed understanding
of the pressure behavior of picene including a classification of all phonon modes,(c)~a~strong indication that doping causes an expansion of the lattice in K-doped samples and (d)~that the lattice
expansion is not the main origin of the frequency softening of the $a_1$~Raman peak observed in K-doped samples.

The paper is organized as follows: in section~\ref{sect:methods}, we give an overview of
the experimental and computational methods; these are described in detail in appendix~A.
In section~\ref{sect:results}, we present the main results of our study, {\em i.e.}
the measured IR and Raman spectra under pressure with a complete assignment of of the most intense 
lines. The {\em ab-initio} equation of state (EOS) of picene is also presented in this section. 
In section~\ref{sect:discussion}, based on the experimental
and first-principles results, we discuss the pressure evolution 
of the vibrational properties of crystalline picene.
We start from the evolution of selected modes with pressure, 
introduce two compact quantities that
can be used to characterise the gradual breakdown of the vibrational picture,
and finally discuss the $a_1$~Raman peak at 1380~cm$^{-1}$.~\cite{Mitamura2011_perpestive-picene,Kambe_2013}
Section~\ref{sect:details} contains the details of the spectral analysis and 
of the applied theoretical concepts, which we believe 
to be useful for further studies.
The main conclusions of our study are summarised in Sect.~\ref{sect:conclusion}.

\section{Methods}
\label{sect:methods}
Solid picene was prepared by a new optimized synthesis route which permits us to obtain sizeable quantities of pure polycrystalline picene powder \cite{Malavasi2011_picene-prep}. Samples have been fully characterized by X-ray diffraction and ambient pressure Raman and IR spectroscopy\cite{JPCM2012_IR_Ram_picene}. 
High pressure Raman and IR measurements have been performed by using diamond anvil cells (DAC). Raman spectra have been collected by using two different instruments with different excitation lines. The first spectrometer  at the Department of Physics of the Sapienza University of Rome was equipped with a He-Ne laser ($\lambda=632.8$~nm), the second one at Porto Conte Ricerche laboratory (Alghero, Italy) was equipped with a diode laser ($\lambda=785$~nm). 
High-pressure IR transmittance data of the picene samples in the DAC were collected at room-temperature exploiting the high brilliance of SISSI beamline of the ELETTRA synchrotron (Trieste, Italy) \cite{Lupi2007_SISSI}. 

For first-principles calculations of the Raman and IR cross sections under pressure, we employed 
Density Functional Perturbation Theory (DFPT),~\cite{Baroni2001_DFPT,Lazzeri2003_Raman_DFT} as implemented in 
the {\em quantum-espresso} code.~\cite{qe} We used standard LDA norm-conserving pseudopotentials.~\cite{psps} 
Further details about experimental and computational procedures are given in the appendix~A.

\section{Results}
\label{sect:results}

\subsection{IR and Raman spectra}
\label{subsect:spectra}

Experimental data and theoretical calculations about Raman and IR spectra at ambient pressure are already available in the literature\cite{JPCM2012_IR_Ram_picene,Girlando2012_IR_Ram_picene}. The picene molecule has C$_{2v}$ symmetry; of the 102 optical modes,
those with  $a_1$~(35), $b_2$~(16), $b_1$~(34) symmetry are both Raman and IR active, while the $a_2$~(17) are only Raman-active~\cite{table:web}. A detailed list of peak frequency, symmetry, optical activity and normalized intensity of the vibrational modes at zero pressure is given in our earlier work \cite{JPCM2012_IR_Ram_picene}. We want to remark that for an isolated molecule, the Raman response is dominated by the (totally symmetric) $a_1$~modes, whereas the IR response is given by the $b_2$~modes \cite{JPCM2012_IR_Ram_picene,Girlando2012_IR_Ram_picene}. Apart from the specific assignment, vibrations with a frequency in the range {200 - 1000}~cm$^{-1}$ can be ascribed to both, out-of-plane and in-plane vibrational modes of carbon and hydrogen, whereas those above 1000~cm$^{-1}$ are basically due to in-plane vibrational modes. Finally modes above 2900~cm$^{-1}$ (not shown here) can be attributed to hydrogen vibrations~\cite{Subedi2011_ele-ph}.

\begin{figure}
\includegraphics[width=8.5cm]{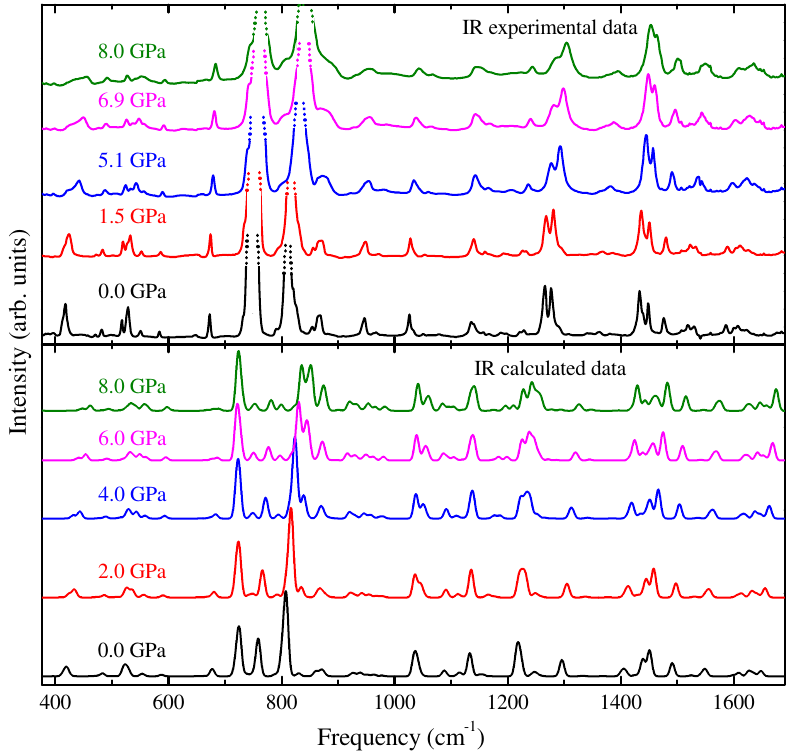}
\caption{(Color online) Experimental (upper panel) and theoretical (lower panel) IR spectra within the range of 400 - 1700~cm$^{-1}$ at selected pressures. We make a convolution of the computed DFT spectra with a Lorentzian profile with 10~cm$^{-1}$ linewidth to ease comparison with experiment. Notice that experimental peaks close to 800~cm$^{-1}$ show saturation effects (look at the dotted lines).}
\label{figs:2} 
\end{figure}

\begin{figure}
\includegraphics[width=8.5cm]{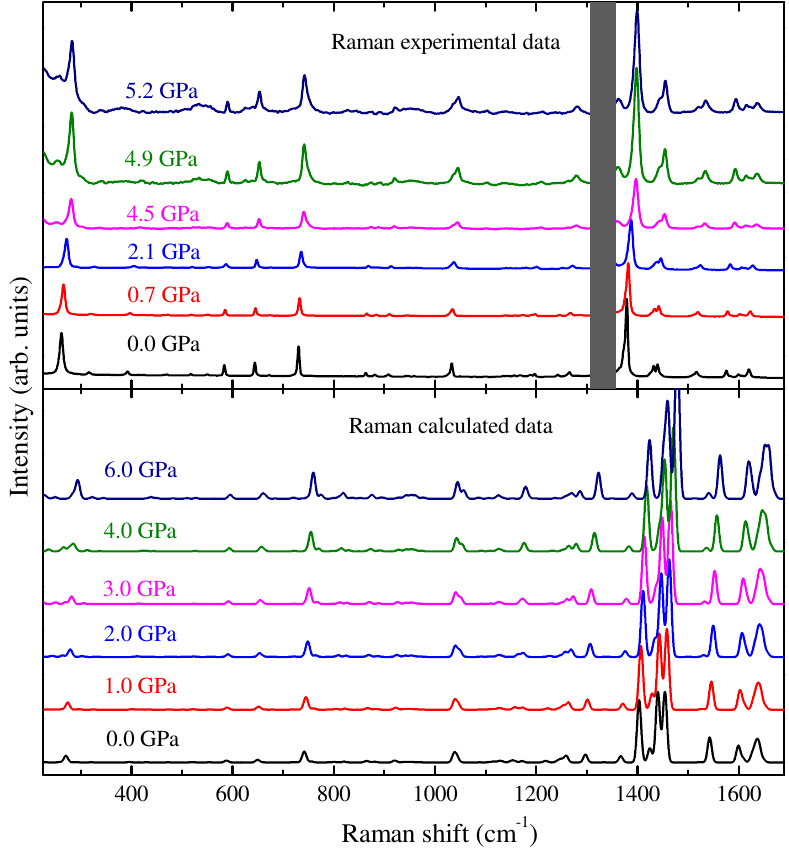}
\caption{(Color online) Experimental (upper panel) and theoretical (lower panel) Raman spectra within the range 200 - 1600~cm$^{-1}$ at selected pressures. We make a convolution of the computed DFT spectra
with a Lorentzian profile with 10~cm$^{-1}$ linewidth to ease comparison with experiment. The shaded region around 1350~cm$^{-1}$ in the experimental spectra is where the Raman peaks of diamond appear.}
\label{figs:3}
\end{figure}

A comparison of the measured and calculated IR and Raman spectra under pressure is shown in Fig.~\ref{figs:2} and \ref{figs:3}. As evident from these figures, all the vibrational  modes display a smooth and uniform hardening with pressure, without any evidence of structural phase transitions. In the experimental IR spectra (Fig.~\ref{figs:2}, upper panel), we notice that the most intense absorption lines result from vanishingly small transmitted intensities and, therefore, can be affected by rather large uncertainties. A good signal to noise ratio is instead obtained for a large number of lower intensity absorption peaks. In the experimental Raman spectra (Fig.~\ref{figs:3}, upper panel) the total integral of each spectrum has been normalized to the sum of the theoretical Raman cross-sections in order to obtain a meaningful comparison between the experimental and theoretical intensities.

A good agreement between experiment and theory is found for both IR and Raman spectra. Calculations reproduce remarkably well the absolute peak frequencies and intensities, including their detailed pressure dependence, over the whole spectral range. The only exception is the high frequency region of the Raman spectra where the calculated peaks intensities appear higher than the experimental ones.
Both calculated and experimental IR spectra show the most intense peaks in the frequency region around 800~cm$^{-1}$. Even the counterintuitive enhancement under pressure of the spectral weight in the 1600~cm$^{-1}$ region in the IR spectra is well reproduced by calculations.
A good example is the IR spectral structure around 1030~cm$^{-1}$ (see Fig.~\ref{figs:2}). On increasing the pressure above 2~GPa the theoretical spectra show a clear splitting of the low-pressure single peak into two components, whereas the experimental spectra show a peak broadening and the evolution toward an asymmetric line shape. It is clear that with the help of the DFT calculation we can cope with a lower experimental definition due to the spectral resolution, pressure gradients, sample inhomogeneities $etc$, and obtain a fine spectral deconvolution. Although less evident, a similar effect is also observed in the Raman spectrum around 730~cm$^{-1}$.
The overall agreement shows that DFT calculations are very reliable in describing this system under lattice compression (see Fig.~\ref{figs:3}). This makes us confident about using the microscopical insight provided by the calculations for a deeper analysis of experimental spectra.  A complete assignment of the vibrational modes between 400 and 1700~cm$^{-1}$ is thus possible.

The frequency, the symmetry and the IR/Raman activity\cite{note:tab:crit} computed for the peaks within the range 700-1700 cm$^{-1}$ are listed in table~\ref{tab:main}. The experimental peak frequencies, as obtained through a detailed lineshape analysis, are also reported (examples can be found in Section~\ref{sect:details:exp} and in appendix B).
Wherever the assignment is unambiguous, we also report the Davydov partners of the main lines~\cite{Davydov}(marked by a bracket ``$\langle$'' in table~\ref{tab:main}). 
Two crystalline modes form a (generalized) Davydov pair if their phonon eigenvector corresponds to the in- and out-of-phase  
linear combination of the same molecular vibration; section~\ref{sect:details:sym} describes in detail how these 
are obtained.
However, we notice that in picene the intensities for Davydov pairs differ by at least an order of magnitude with an exception of the IR-active pair around 800~cm$^{-1}$,
therefore Davydov splittings are hardly accessible experimentally (see also Fig.~\ref{figs:8}). 

\begin{table}[tbp]
\begin{ruledtabular}
  \begin{tabular}{r@{ }rccr@{ }lr@{ }lrr}
     \multicolumn{2}{c}{$\nu_{\mathrm{DFT}}$} & Sym & \multicolumn{3}{c}{$I_{\mathrm{IR}}$} & \multicolumn{2}{c}{$I_{\mathrm{R}}$} & \multicolumn{1}{c}{$\gamma_{\mathrm{DFT}}$} & \multicolumn{1}{c}{$\nu_{\mathrm{exp}}$} \rule[-1.2ex]{0pt}{0pt}\rule{0pt}{2.6ex}\\
     \multicolumn{2}{c}{(cm$^{-1}$)} & & \multicolumn{3}{c}{({\footnotesize $(\mathrm{D}/\mathrm{\AA})^2/ \mathrm{amu}$})} & \multicolumn{2}{c}{ ({\footnotesize $\mathrm{\AA}^4 / \mathrm{amu}$})} & & (cm$^{-1}$)\\
    \hline
                                 &  725.6 & $b_2$ & & \textbf{5.24} & ($\downarrow$) &             2  & ($\uparrow$)   &   0.00 & 740\rule{0pt}{2.6ex} \\
\multirow{2}{*}{$\Big\langle$}   &  756.4 & $b_2$         & &         0.89  & ($\downarrow$) &            78  & ($\uparrow$)   &   0.14 &   -  \\
                                 &  759.6 & $b_2$         & & \textbf{3.67} & ($\downarrow$) &             0  & ($\uparrow$)   &   0.12 &  756\\

\multirow{2}{*}{$\Big\langle$}   &  801.3 & $b_2$         & & \textbf{2.68} & ($\downarrow$) &            64  & ($\uparrow$)   &   0.13 & \multirow{2}{*}{$\Big\rbrace\quad 810$}\\
								 &  808.6 & $b_2$         & & \textbf{9.75} & ($\downarrow$) &             2  & ($\uparrow$)   &   0.13 &     \\
\multirow{2}{*}{$\Big\langle$}   & 1035.1 & $a_1$ + $b_1$ & & \textbf{2.33} & ($\uparrow$)   &            11  & ($\uparrow$)   &   0.02 & 1025\\
  								 & 1037.7 & $a_1$ + $b_1$ & &         0.07  & ($\downarrow$) &           938  & ($\uparrow$)   &   0.04 &  -  \\
							     & 1133.9 & $b_1$         & & \textbf{1.55} & ($\uparrow$)   &            25  & ($\downarrow$) &   0.04 & 1134\\
								 & 1216.4 & $b_1$         & & \textbf{1.72} & ($\uparrow$)   &             7  & ($\uparrow$)   &   0.04 & 1265\\
								 & 1223.8 & $a_1$         & & \textbf{1.64} & ($\uparrow$)   &             5  & ($\uparrow$)   &   0.10 & 1276\\
\multirow{2}{*}{$\Big\langle$}   & 1400.4 & $a_1$         & &         0.15  & ($\uparrow$)   &            22  & ($\downarrow$) &   0.06 &   - \\
                                 & 1403.1 & $a_1$         & &         0.47  & ($\uparrow$)   &  \textbf{6829} & ($\uparrow$)   &   0.08 & 1377\\
                                 & 1424.1 & $a_1$         & &         0.00  &                &  \textbf{1482} & ($\downarrow$) &   0.10 & 1433\\
\multirow{2}{*}{$\Big\langle$}   & 1438.1 & $a_1$         & &         0.77  & ($\downarrow$) &            18  & ($\uparrow$)   &   0.08 &   - \\
                                 & 1440.0 & $a_1$         & &         1.25  & ($\downarrow$) &  \textbf{7937} & ($\downarrow$) &   0.08 & 1433\\
        						 & 1451.0 & $b_1$         & & \textbf{3.05} & ($\uparrow$)   &            14  & ($\downarrow$) &   0.08 & 1450\\
                                 & 1451.8 & $b_1$         & &         0.00  & ($\downarrow$) &  \textbf{1804} & ($\uparrow$)   &   0.08 & 1440\\
\multirow{2}{*}{$\Big\langle$}   & 1454.2 & $a_1$         & &         0.02  & ($\uparrow$)   &           108  & ($\uparrow$)   &   0.08 &   - \\
                                 & 1454.5 & $a_1$         & &         0.03  & ($\uparrow$)   &  \textbf{6309} & ($\uparrow$)   &   0.09 & 1440\\
\multirow{2}{*}{$\Big\langle$}   & 1491.3 & $b_1$         & & \textbf{1.59} & ($\uparrow$)   &             1  & ($\downarrow$) &   0.07 & 1475\\
							     & 1493.6 & $b_1$         & &         0.01  & ($\uparrow$)   &             1  & ($\uparrow$)   &   0.07 &   - \\
\multirow{2}{*}{$\Big\langle$}   & 1542.3 & $a_1$         & &         0.05  & ($\uparrow$)   &  \textbf{2792} & ($\uparrow$)   &   0.07 & 1516\\
                                 & 1544.5 & $a_1$         & &         0.28  & ($\uparrow$)   &             5  & ($\downarrow$) &   0.07 &   - \\
\multirow{2}{*}{$\Big\langle$}   & 1599.0 & $b_1$         & &         0.12  & ($\downarrow$) &  \textbf{1518} & ($\downarrow$) &   0.07 & 1574\\
                                 & 1599.1 & $b_1$         & &         0.00  & ($\uparrow$)   &           380  & ($\uparrow$)   &   0.06 &   - \\
                                 & 1633.3 & $a_1$         & &         0.18  & ($\uparrow$)   &  \textbf{1714} & ($\uparrow$)   &   0.06 & \multirow{2}{*}{$\Big\rbrace\quad 1620$}\\
                                 & 1639.3 & $a_1$         & &         0.04  & ($\uparrow$)   &  \textbf{1732} & ($\uparrow$)   &   0.07 &    \\
  \end{tabular}
\end{ruledtabular}
  \caption{Assignment of the most intense phonon peaks with their  Davydov partners, if existing.
   Calculated and experimental phonon frequencies are reported in column $\nu_{\mathrm{DFT}}$ and $\nu_{\mathrm{exp}}$.  
      The mode symmetry of the largest molecular
           component of the crystal eigenstate is given in the second column. $I_{\mathrm{IR}}$ and $I_{\mathrm{R}}$ are calculated IR and Raman intensities;
           $\uparrow$ and $\downarrow$ indicate that the intensity increases/decreases with pressure.
DFT Gr\"uneisen parameters, calculated according to  Eq.~(\ref{eq:gr}), are reported in column $\gamma_{DFT}$. The brackets ``$\langle$'' mark the two members of Davydov pairs.}
   \label{tab:main}
\end{table}

Table~\ref{tab:main} also reports the values of the 
 the Gr\"uneisen parameters $\gamma$ \cite{Postorino_2002} of the vibrational modes of solid picene evaluated according to the formula:
\begin{equation}
  \gamma = - \frac{\Delta\nu / \nu_0}{\Delta V / V_0}
  \label{eq:gr}
\end{equation}
Since the pressure versus volume ($p$ vs $V$) relation was not accessible experimentally, we used the theoretical equation of state (EOS) (see sect.~\ref{sect:eos}) to determine $\Delta V$ in eq.~\ref{eq:gr}. 
For all modes, both 
 calculations and experimental results consistently show positive $\gamma$ values, {\em i.e.}~all frequencies {\em harden} upon lattice compression.
The values reported in the table ({$\gamma_{DFT}$}) are those extracted from DFT calculations  which however
agree very well with our measured data.

It is particularly interesting to compare our pressure results with two different 
 phonon measurements in K doped samples, performed by the same group --
see Ref.~\mbox{[\citenum{Mitamura2011_perpestive-picene}] and [\citenum{Kambe_2013}]}.
 Both these works report on a frequency softening with increasing K doping, whereas opposite effects are claimed for the volume: a compression is reported in Ref.~\mbox{[\citenum{Mitamura2011_perpestive-picene}]}, an expansion in Ref.~\mbox{[\citenum{Kambe_2013}]}. Our data clearly support the latter picture. In fact, a  close  inspection  of  the  spectra  reported  in Ref.~\mbox{[\citenum{Mitamura2011_perpestive-picene}]} shows a frequency softening of all the observed Raman peaks on increasing the K content. A general softening clearly indicates a lattice expansion. Other 
effects that could lead to the softening  of phonon frequencies in K doped samples,
such as  electron-phonon coupling or charge transfer, would be extremely mode-dependent.
In section~\ref{sect:disc:a1}, we will continue this discussion, focusing on the $a_1$~Raman-active mode at 1380~cm$^{-1}$, which
has been used as a marker for electron transfer in alkali metal intercalated picene~\cite{Mitamura2011_perpestive-picene, Kambe_2013}.

\subsection{Equation of State}
\label{sect:eos}
We already mentioned above that the $p$ vs $V$ relation was experimentally not accessible. 
Therefore, for the EOS, we have to resort to our {\em ab-initio} (DFT)
calculations. 
In order to maintain a reasonable balance between accuracy and computational cost, we employed the local density approximation (LDA). 
It is well-known that LDA leads to an overbinding in van der Waals (vdW) solids, but, compared to other widely-used
functionals (GGA), it predicts in most cases stable structures and yields reasonable values of elastic 
constants and phonon frequencies.~\cite{Hasegawa_2007, Ortmann_2006, Marzari_2005}.

\begin{figure}[b]
  \centering
  \includegraphics[width=0.48\textwidth]{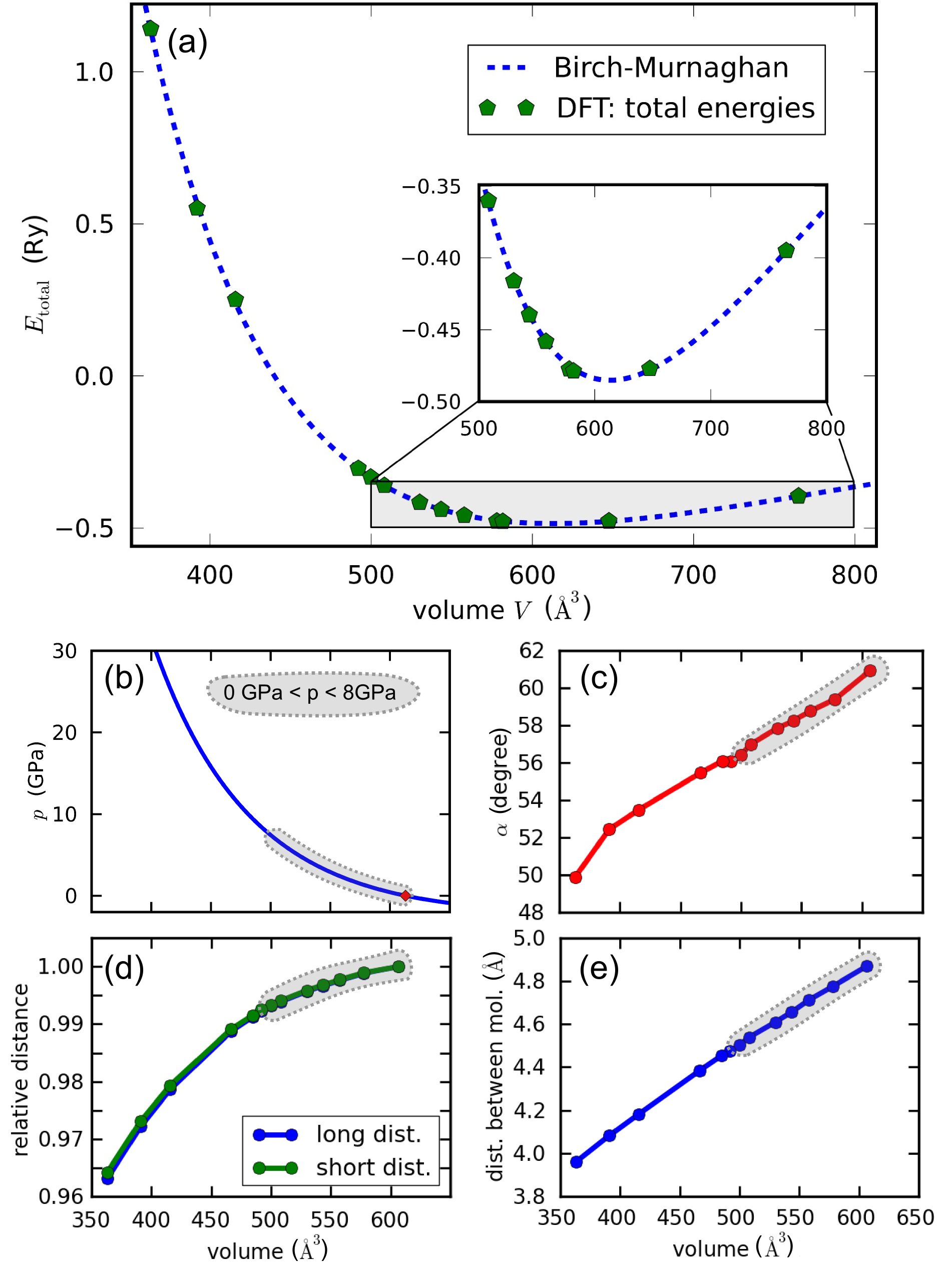}
  \caption{(Color online) Equation of state of solid picene under pressure, from DFT calculations.
           (a)~Energy vs. volume relation. Total energies from density functional theory~(symbols), fitted with Birch-Murnaghan equation of state~(line). The inset shows an enlargement of the region around the energy minimum. (b)~Corresponding $p$~vs~$V$ relation. (c)~Evolution of the intermolecular angle $\alpha$; (d)~intra-molecular long and short distances $ld$ and $sd$ divided by their equilibrium value -- for definitions, see Fig.~\ref{figs:1}; (e)~distance between the geometric centres of the two molecules, as a function of the unit-cell volume. 
} %
  \label{figs:4}
\end{figure}

We computed the total energy as a function of volume, for fully-optimised unit cells between 50\% and 110\% of
 the experimental volume.
Fig.~\ref{figs:4} (a) shows the results of our calculations for the Equation of State~(symbols), 
together with a fit to the Birch-Murnaghan relation~(line).
We obtained an equilibrium volume $V_0 = 613$~\AA$^3$, which is 13\% smaller than the experimental value\cite{Kambe_2013} \mbox{$V_{\mathrm{exp}} = 708$~\AA$^3$}, and a bulk modulus $B_0 = 18.5$~GPa with a derivative $B'_0 = 6.8$~\cite{foot:vdw}.
The bulk modulus is the same order of magnitude as measured in a recent study for phenanthrene under pressure~\cite{phena2013}.
The authors also report a $B'_0$ value anomalously small, roughly four times smaller than ours. The discrepancy may be due to the fact that the samples in Ref.~[\onlinecite{phena2013}] show a coexistence of \textit{P2/m} and \textit{Pmmm} phases. 

It is worth to notice that our calculated compressibility for crystalline picene is typical for a molecular system. Such a value naturally implies a large sensitivity of phonon modes with pressure, and indeed we find a remarkable hardening for all modes. The 8~GPa range accessible to our IR and Raman measurements corresponds to a 20$\%$~reduction of the unit-cell volume with respect to its equilibrium value. This range is shown as a grey-shaded area in Fig.~\ref{figs:4}~(a)-(e). 

Besides a decrease of the unit cell volume, pressure also causes a change in the relative coordinates of the two molecules inside the unit cell.
In particular, the angle $\alpha$ between the molecules decreases monotonically,
favouring a more parallel alignment -- see Fig.~\ref{figs:4}~(c). As a consequence, the distance between the two geometrical centres shrinks -- see Fig.~\ref{figs:4}~(e). However, the contraction of the C$_{22}$H$_{14}$ molecules themselves, measured by the  two intramolecular distances $sd$ and $ld$, is much smaller, {\em i.e.} less than 1$\%$ over 8~GPa -- see Fig.~\ref{figs:4}~(d). We also observe a slight bending of the molecules, which increases with pressure. We do not observe any structural transition, in accordance with experiment (see Sect.\ref{subsect:spectra}).
Moreover, it has been observed that picene remains insulating at room temperature up to~25~GPa;
\cite{Fanetti2013_HP} in the same pressure range, present calculations show that the LDA band gap decreases from $\Delta \approx 2$~eV to $\Delta \approx 0.8$~eV. 

\section{Discussion}
\label{sect:discussion}

The detailed comparison of experiments and calculations allows us to characterize the behaviour of crystalline picene under pressure. 
In particular, the comprehensive knowledge of the phonon eigenvectors in the DFPT calculations
not only allowed us to track the evolution of individual modes under pressure,
but also to follow the gradual increase in intermolecular interactions. 

\subsection{Phonon Modes under Pressure}

In the crystalline form of picene intermolecular interactions 
modify the phonon spectrum with respect to that of the
pure molecule. This effect increases with pressure and,
although impossible to access experimentally, it can 
be easily quantified with the help of first-principles calculations,
using projections on molecular eigenvectors, as described in 
Sect.~\ref{sect:details:sym}.

In the spectrum, we can identify three types of modes:
\begin{itemize}
\item Modes which have a definite molecular character at $p$=0,
and retain it up to high pressures ({\em molecular} modes).
\item Modes which have a definite molecular character at $p$=0,
but lose it with pressure ({\em mixed} modes).
\item Modes which already at $p$=0 have no definite molecular character
({\em crystalline} modes).
\end{itemize}

We can distinguish between these modes introducing 
the maximum projection on a molecular eigenmode, defined as 
$\Pi_j = \mathrm{max}_i \left| \langle \psi_i | \Psi_j \rangle \right|^2$, where $\psi_i$ are the eigenvectors of the single molecule and $\Psi_j$ are the corresponding eigenvectors in the solid (for a detailed description see Sect.~\ref{sect:details:sym}). 
Using this quantity, \textit{molecular} modes are defined as modes which have $\Pi \geq 0.9$, and retain a
$\Pi \geq 0.8$ at $p=6$~GPa.  \textit{Crystalline} modes are those
which at zero pressure have $\Pi < 0.9$,and thus cannot be represented as a single product
of molecular vibrations at any pressure.
Notice that this classification, being based on a quantitative definition in terms
of partial eigenvectors, is different from the usual one in terms of intra- and inter-molecular
phonons. In particular, crystalline modes are not simply rigid translations of 
vibrations of the full molecule, and thus occur also at large frequencies.

Crystalline modes represent $\sim 1/4$ of the total number of modes at $p=0$~GPa;
molecular modes represent only $\sim 1/10$, and the rest 
is represented by mixed modes.
However, molecular modes have generally a high intensity, according to 
 their Raman and IR selection rules.

\begin{figure}
\includegraphics[width=8.5cm]{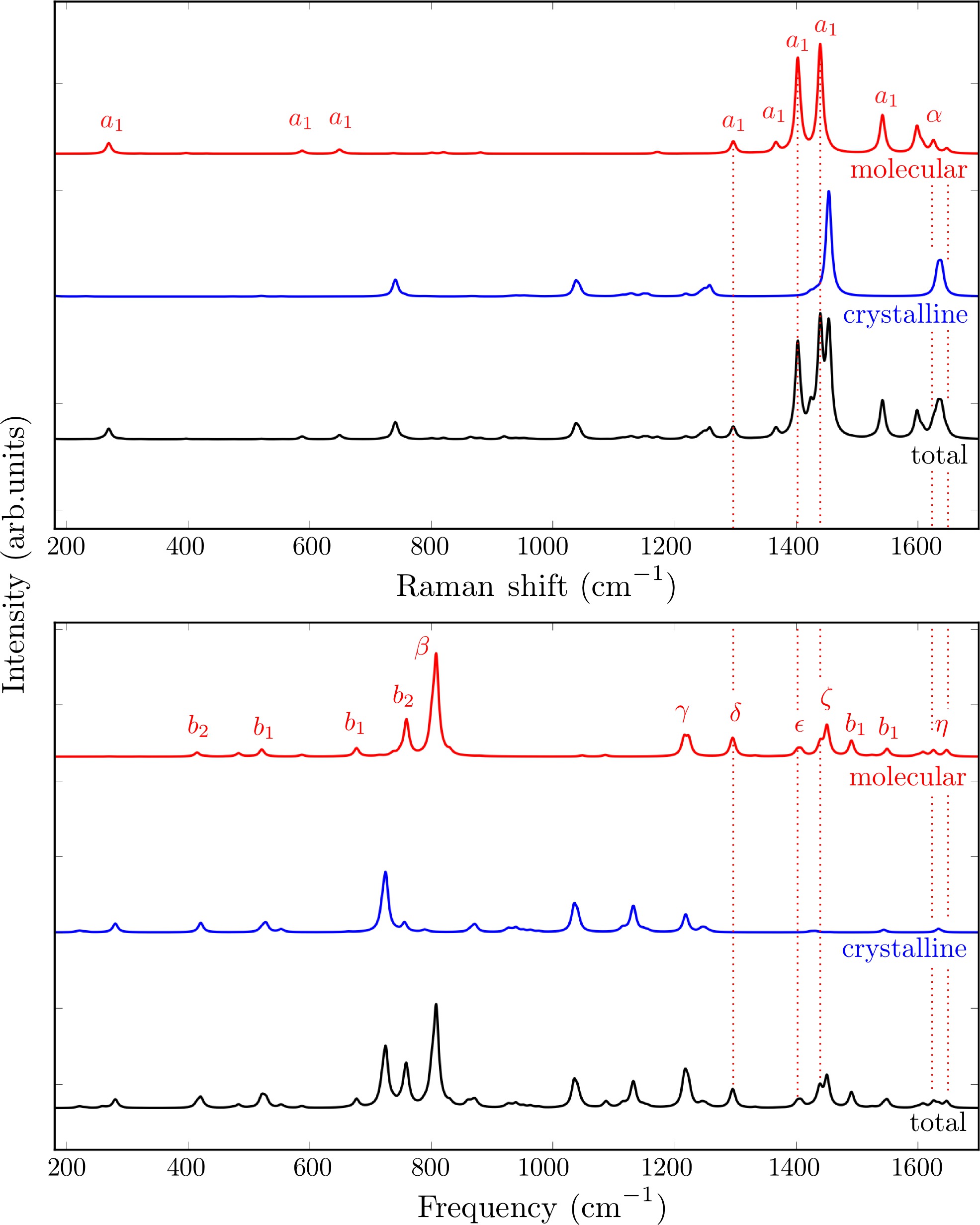}
\caption{(Color online) Decomposition of the theoretical Raman (top) and
IR (bottom) spectra, into crystalline and molecular modes (see text).
The irreducible representation
 of the corresponding molecular modes is indicated in correspondence
of the peaks; greek letters indicate peaks that result from the superposition
of more than one mode (see table~\ref{tab:press}).
The red-dotted lines are guidelines for the eyes for modes which are both IR and Raman active.}
\label{figs:5}
\end{figure}

Fig.~\ref{figs:5} 
illustrates the relative importance of molecular and crystalline modes on the IR and Raman spectra,
using the DFT spectra.
The total spectrum is shown in black, molecular and crystalline modes are plotted in
red and blue respectively. We notice that
high-intensity crystalline peaks appear at $\sim 1400$ and $\sim 1600$~cm$^{-1}$ in the Raman 
spectra, at $\sim 700$~cm$^{-1}$ in the IR, and several other peaks should be clearly
resolvable in experiment.

On the figure, we also indicated the corresponding irreducible representation of the most intense
molecular peaks.
Where the assignment is problematic, because several almost degenerate peaks occur 
at the same energy, we use greek letters. 
The frequencies are then reported in table~\ref{tab:press}, which is a collection
of purely \textit{molecular} modes with a finite contribution to the IR/Raman~spectrum.
In the table, out-of-plane modes are indicated with a star ($\star$). We notice
that there is no particular predominance of in- or out-of-plane character
in the molecular modes.
The existence and relative weight of crystalline 
and mixed modes can be used to quantify the intermolecular interaction, as
illustrated in the following. 
\begin{table}
 \begin{ruledtabular}
 \begin{tabular}[b]{r@{ }rrrc}
 &\multicolumn{1}{c}{$\nu_{\mathrm{DFT}}$} & \multicolumn{1}{c}{$\Pi$} & \multicolumn{1}{c}{Irr.} & \multicolumn{1}{c}{opt.}\rule{0pt}{2.6ex}\\
 &\multicolumn{1}{c}{(cm$^{-1}$)} & & Rep.  & act.\rule{0pt}{2.6ex}\\
\hline
             & \textbf{269.8} & 0.91	& $a_1$         & R\rule{0pt}{2.6ex}\\
             & \textbf{415.1} & 0.98	& $b_2$ $\star$ & IR\\
             & \textbf{521.5} & 0.96	& $b_1$         & IR\\
             & \textbf{587.9} & 0.99	& $a_1$         & R\\
             & \textbf{649.0} & 0.99	& $a_1$         & R\\
             & \textbf{677.3} & 0.98	& $b_1$         & IR\\
             &  759.6 & 0.94	& $b_2$ $\star$ & IR\\
$\beta$       &  801.3 & 0.96	& $b_2$ $\star$ & IR\\
$\beta$       &  808.6 & 0.97	& $b_2$ $\star$ & IR\\
$\beta$       & \textbf{830.8} & 0.91	& $b_2$ $\star$ & IR\\
$\gamma$      & 1216.4 & 0.96	& $b_1$         & IR\\
$\gamma$      & 1223.8 & 0.92	& $a_1$         & IR \\
$\delta$      & \textbf{1294.8} & 0.99	& $a_1$         & IR\\
$\delta$      & \textbf{1297.0} & 0.98	& $a_1$         & IR/R\\
              & \textbf{1367.0} & 0.99	& $a_1$         & R\\
$\epsilon$    & 1403.1 & 0.97	& $a_1$         & IR/R\\
$\epsilon$    & \textbf{1408.7} & 0.96	& $b_1$         & IR\\
$\zeta$       & 1440.0 & 0.93	& $a_1$         & IR/R\\
$\zeta$       & 1451.0 & 0.93	& $b_1$         & IR\\
             & 1491.3 & 0.99	& $b_1$         & IR\\
             & 1542.3 & 0.91	& $a_1$         & R\\
             & \textbf{1549.9} & 0.93	& $b_1$         & IR\\
$\alpha$      & 1599.0 & 0.98	& $b_1$         & R\\
$\alpha$      & \textbf{1599.1} & 0.97	& $b_1$         & R\\
$\eta,\alpha$ & \textbf{1607.8} & 0.95	& $a_1$         & R\\
$\alpha$      & \textbf{1626.1} & 0.96	& $b_1$         & R\\
$\eta,\alpha$ & \textbf{1626.1} & 0.96	& $b_1$         & IR/R\\
$\eta,\alpha$ & \textbf{1647.8} & 0.97	& $b_1$         & IR/R\\
 \end{tabular}
 \end{ruledtabular}
 \caption{\label{tab:press}
Calculated phonon frequencies $\nu_{\mathrm{DFT}}$ at $p=0$~GPa, which have a molecular character $\Pi_j~=~\mathrm{max}_i \left|  \langle \psi_i | \Psi_j \rangle \right|^2 \geq 0.9$ and a finite IR / Raman cross section.
Bold modes do not appear in table~\ref{tab:main}, due to small
cross-section or energy range.
 The irreducible representation of the corresponding molecular mode is given in the fourth column. The star ($\star$) next to the irreducible  representation marks modes which have mainly out-of-plane character; 
the others are predominantly in-plane.} 
\end{table}

\subsection{Intermolecular Interactions}
\begin{figure}
  \centering
  \includegraphics[width=8cm]{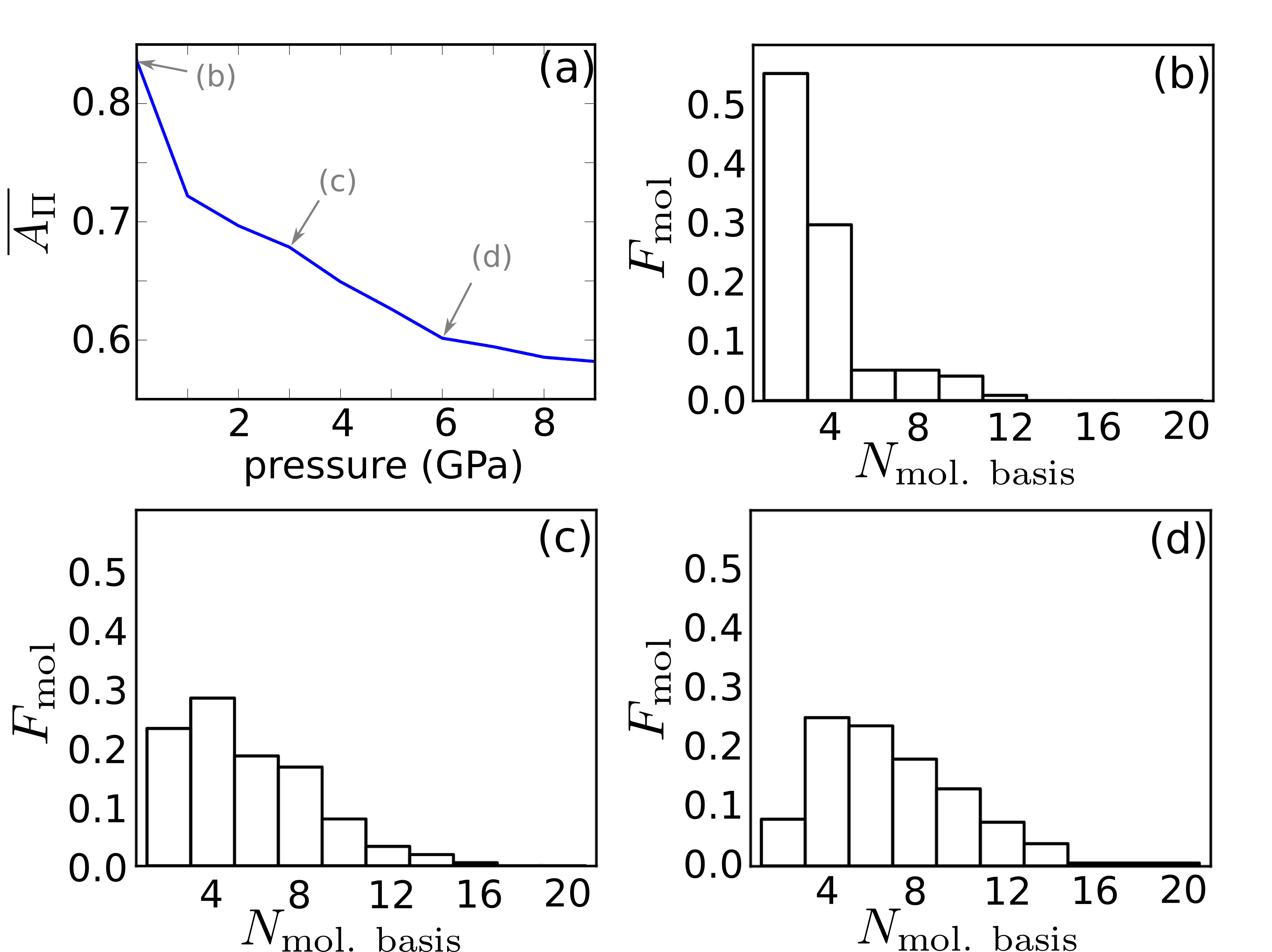}
  \caption{(Color online) (a) The mean of the largest projections
$\overline{A_{\mathrm{\Pi}}}$
  vs. pressure. (b-d) Relative frequency of the molecular basis states $F_{\mathrm{mol}}$ needed to reconstruct the crystal phonon mode with an accuracy of at least 0.9 for $p=(0,3,6)$~GPa.}
  \label{figs:6}
\end{figure}

In order to obtain a {\em compact} estimate of the loss of molecular character with pressure,
we introduced two different quantities.
These are plotted in Fig.~\ref{figs:6}.
$F_{\mathrm{mol}}$ is the fraction of crystal phonon modes,
whose eigenvectors can be represented as a superposition of $N_{\mathrm{mol}}$ molecular eigenstates
with an accuracy of at least 90~$\%$. For a perfect molecular crystal, $F_{\mathrm{mol}}$ would comprise a single
peak, with weight one, at $N_{\mathrm{mol}}=2$. For $p=[0,3,6]$~GPa, the corresponding histograms are shown in the
panels~$(b)$-$(d)$ of Fig.~\ref{figs:6}.
For solid picene at zero pressure, $F_{\mathrm{mol}}$=0.5 for $N_{\mathrm{mol}}=2$, 0.3 for $N_{mol}=4$, and decays rapidly for increasing $N_{mol}$. Already at ambient conditions, only 50$\%$ of the crystal eigenmodes can be clearly assigned to a given simple product of molecular vibrations. This fraction rapidly decreases with pressure, whereas the weight of states with high $N_{\mathrm{mol}}$ increases. 

A compact measure for the change of the character of phonons with 
pressure is given by $\overline{A_{\mathrm{\Pi}}}$, which is the average
over the largest coefficients squared of the expansion of a single crystalline eigenmode in molecular vibrations (for a detailed definition see the Sect.~\ref{sect:details:sym}).
In case of a molecular crystal, this number is 1, {\em i.e.} all crystalline eigenmodes can be decomposed in a single product state of molecular vibration. In the crystalline limit,
{\em i.e.} equally-distributed projections onto the molecular basis set, we would have in the case of picene:
$\overline{A_{\mathrm{\Pi}}}=1/\left(36\times3\right)=1/108\approx0.01$. For picene, $\overline{A_\mathrm{\Pi}}$ vs. pressure is plotted in Fig.~\ref{figs:6}$(a)$. At ambient pressure, it is still in the vicinity of a molecular crystal ($\overline{A_{\mathrm{\Pi}}}\approx 0.8$). With increasing pressure, $\overline{A_{\mathrm{\Pi}}}$ decreases, but at $p=8$~GPa it is still far from the crystalline limit ($\overline{A_{\mathrm{\Pi}}} = 0.6$). In addition, we did a calculation at $p\approx 35$~GPa, and found that even there, picene remains close to the molecular limit ($\overline{A_{\mathrm{\Pi}}}=0.5$). However, a simple picture of two largely independently vibrating molecules breaks down with increasing pressure.
The loss of molecular character can invalidate simple vibrational models that are used to interpret 
phonon spectra and to evaluate the electron-phonon interaction in solid 
picene.~\cite{Kato2011_ele-ph,Sato2012_ele-ph}
Similar, but more pronounced effects have been discussed for K-doped solids.~\cite{Casula2011_ele-ph} 

In the present case, 
even when there is no one-to-one correspondence to molecular eigenmodes, it is still possible to follow the evolution of crystalline modes with pressure, and define (generalised) Gr\"uneisen parameters and Davydov splittings, as explained in Sect.~\ref{sect:details:sym}.
We have reported them in table~\ref{tab:main}.

\subsection{The $a_1$ Raman peak at 1380~cm$^{-1}$}
\label{sect:disc:a1}
\begin{figure}
  \includegraphics[width=8cm]{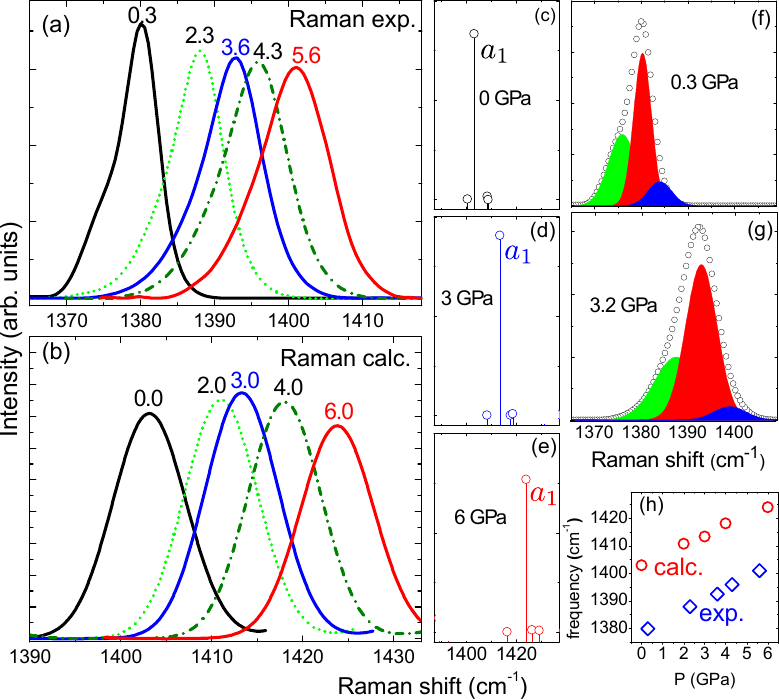}
  \caption{(Color online) The Raman modes of picene around 1380~cm$^{-1}$ at different external pressures. 
  The experimental data are presented in panel (a). The computed spectra presented in panel (b) are obtained 
through a convolution of
the DFT cross sections with a Lorentzian function to mimic 
  the experimental broadening -- see panels (c)-(e). 
Examples of the spectral deconvolution at two pressures are shown in panels (f) and (g). 
  Panel (h) shows the pressure dependence of the frequency of the component which has 
the highest intensity in experiment (blue)  and theory (red). 
  }
  \label{figs:7}
\end{figure}

Here we discuss in detail the $a_1$ Raman peak with $\nu_0 \sim 1380$ cm$^{-1}$ (DFT frequency: 1403 cm$^{-1}$) at ambient pressure.
The peak frequency has been used in Refs.~[\citenum{Mitamura2011_perpestive-picene}] and [\citenum{Kambe_2013}] as a marker for doping in $\mathrm{K_{x}picene}$. 
Moreover, according to Refs.~[\onlinecite{Casula2011_ele-ph,Subedi2011_ele-ph,Kato2011_ele-ph,Sato2012_ele-ph}]  this mode displays the highest coupling to electrons upon K intercalation. 
 
Experimental and theoretical results, reported in Fig.\ref{figs:7}(a) and (b),  clearly show that the spectral structure is dominated by the $a_1$ peak (see Fig.\ref{figs:7}(c)-(e)). The overall spectral shape, under compression, broadens but does not drastically change (see panels (f) and (g)). DFT calculations well reproduce this result (see Fig.\ref{figs:7}(b)) and also the pressure dependence of the peak frequency (see panel (h)). The most intense Raman line retains a well-defined molecular character ($a_1$) up to 6 GPa (the projection is 97 \% at 0~GPa and 86 \% at 5~GPa), and has a low-intensity Davydov partner (see Fig.~\ref{figs:10}).

Focusing on the remarkably strong frequency hardening of the central peak and exploiting both experimental and theoretical data, we find  $\Delta\nu_p\sim$+23~cm$^{-1}$ for a pressure variation of 5.2 GPa. The corresponding variation of the volume of the unit cell, from the DFT data in Fig.~\ref{figs:3}, is $\Delta V_p\sim$100~\AA$^3$, which yields a
 Gr\"{u}neisen parameter $\gamma_p \sim 0.1$. This extremely low value is expected, due to the strongly harmonic character of this high frequency vibrational mode, and it is in good agreement with the values reported in the literature for similar modes in aromatic compounds\cite{Zhao_anthracene_1999} -- Notice also the excellent agreement with the calculated $\gamma$ value  ($\gamma_{\mathrm{DFT}}=0.08$) reported in Table~\ref{tab:main}. 

We can use the  Gr\"{u}neisen parameter $\gamma_p$ to estimate the relative importance of structural and doping (charge transfer, electron-phonon coupling) effects
in K-doped picene. {\em Kambe et al.}~\cite{Kambe_2013} report for $x=3$ a negative shift $-65~cm^{-1}$ of the $a_1$ frequency, with a 6 $\%$ relative volume {\em expansion}
-- see Table~\ref{tab:a1} and section~\ref{sect:results} for more details.

\begin{table}[ht]
\begin{ruledtabular}
\begin{tabular}[b]{c c c c}
& \quad $\Delta V / V_0$ & \quad $\Delta\nu /\nu_0$ & \quad $\gamma$\\
\hline
$\mathrm{K_3 picene}$\cite{Kambe_2013} & \quad +6\% & \quad -5\% & \quad $\gamma_d \sim 0.8$\\
picene at 5.2~GPa & \quad -15\% & \quad +1.6\% & \quad $\gamma_p \sim 0.1$\\
\end{tabular}
\end{ruledtabular}
\caption{Relative volume and frequency variation and Gr\"uneisen parameters of the $a_1$ Raman peak  at 1380~cm$^{-1}$. Data are presented for $\mathrm{K_3}$ doped picene (first row, according to Ref.[\citenum{Kambe_2013}]) and pristine picene for a pressure change of 5.2~GPa (second row, present work).}
\label{tab:a1}
\end{table}

This corresponds to an effective Gr\"uneisen parameter $\gamma_d \sim 0.8$, which is much larger than the purely structural value $\gamma_p \sim 0.1$.
This is a strong indication that in K-doped samples factors other than structure, such as charge doping or electron-phonon coupling,
determine the softening of the $a_1$ mode.

\section{Details of Data Analysis}
\label{sect:details}

The analysis of the pressure evolution of the Raman and IR spectra of a system which comprises $2 \times 3 \times 36=216$ phonon modes, most of which are Raman and IR active, is a formidable task.
Besides the obvious difficulty of assigning a large number of phonon peaks, which are often almost degenerate in energy, there is the additional complication that, with increasing pressure, 
the shape of the spectra changes not only because of the hardening or softening of the phonon modes, but also because of the changes in the relative optical cross sections, due to the rearrangement of the internal coordinates.
When the number of optically active modes is large, these two effects are impossible to disentangle in the experimental spectra, and can lead to severe misinterpretation of the data. 
Therefore, using the information from the corresponding DFPT~calculations, we can achieve a very detailed lineshape analysis of the measured spectra. 
We provide here one example of this analysis, relative to the 
IR spectra in the 400-480~cm$^{-1}$ range.
The appendix~B contains additional examples of 
deconvolution of spectral structures, in particular Raman in the 570-790~cm$^{-1}$ spectral range 
and IR in the 1390-1500~cm$^{-1}$ range.

\subsection{Deconvolution of Complex Spectral Features}
\label{sect:details:exp}
\begin{figure}
\includegraphics[width=8cm]{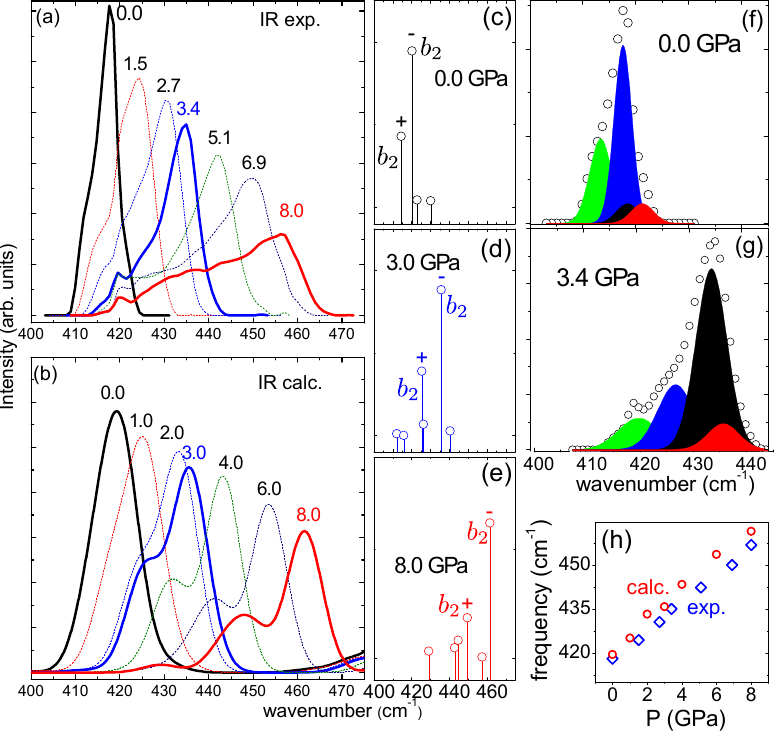}
\caption{(Color online) The IR modes of picene around 415~cm$^{-1}$ at different external pressures.
Panels are organized according to the same scheme as in Fig.\ref{figs:7}. 
In panels  (c)-(e), we indicate the symmetry of the main molecular modes
for the most intense calculated peaks.
With $+/-$ we indicate the two components of a Davydov pair.}
\label{figs:8}
\end{figure}

An example of deconvolution of complex spectral features is
shown in Fig.~\ref{figs:8}, where the experimental IR spectra
are compared with the linear response calculations over the 400-480~cm$^{-1}$ frequency range.
Experimental data are shown in panel (a); here, it appears
that the band centered around 415~cm$^{-1}$ at zero pressure evolves towards a broad and strongly asymmetric structure. 
The DFT calculations (see Fig.~\ref{figs:8}(b)) reproduce remarkably 
well the evolution of the experimental spectrum.
At 0.0~GPa, the excellent agreement between calculated (Fig.~\ref{figs:8}(c)) and measured spectra (Fig.~\ref{figs:8}(f)) allowed us to 
resolve four different components, capturing both peak positions and intensities. At 3.4~GPa, a clear line broadening is observed (see Fig.~\ref{figs:8}(g)).
Calculated spectra clearly show that the effect of pressure is to enhance a few peaks 
that have vanishingly small intensity at lower pressure. This is quite clear looking at Fig.~\ref{figs:8}(d) for $p = 3$ GPa and even more in Fig.~\ref{figs:8}(e) for $p = 8$ GPa. 
We can thus argue that the broadening of the spectra reflects the ``appearance" of additional modes, which are hardly detectable at zero pressure. The  pressure evolution of the most intense component, which has $b_2$ character at zero pressure, is shown in Fig.~\ref{figs:8}(h). The agreement between calculations and experiment is extremely good.

In Fig.~\ref{figs:8}(c) we can also notice that the two most intense peaks, indicated with $+$ 
and $-$, form a Davydov pair. In general, the evolution of the splitting of the frequencies of a Davydov pair is a good measure for increasing intermolecular interactions in a crystal. Nevertheless, in this case, on increasing the pressure (see panels (d) and (e)) several modes acquire a finite spectral weight and cover the Davydov pair, thus making it impossible to disentagle the position of the Davydov pair without the information from {\em ab-initio} calculations.
Possible alternative indicators of intermolecular interactions, which should be more suitable for large systems, have been discussed in the previous section.

\subsection{Symmetry Analysis of the Phonon Modes}
\label{sect:details:sym}

At ambient pressure, the main features of the Raman and IR spectra of solid picene are captured by a simplified vibrational analysis.~\cite{Kato2011_ele-ph,Sato2012_ele-ph,Girlando2012_IR_Ram_picene,Kambe_2013} One essentially assumes that there is 
 a one to two correspondence between the vibrational eigenvalues and eigenvectors ($\nu^{\alpha}_{n}$ and $\psi^{\alpha}_n$; $\alpha$= $a,b$; $n$=$1,,...,108$) of an isolated picene molecule,
and the corresponding phonon frequencies and eigenvectors ($\mathrm{N}_{m}$, $\Psi_m$, $m$=$1,...,216$) in the solid. 
Each phonon in the crystal can be identified with one of the two modes, in which two molecules $\alpha=a,b$ vibrate in or out-of-phase with each other, according to a given eigenvector $n$. The small energy difference between the two modes (Davydov splitting) is often used as a measure for intermolecular interaction.~\cite{Davydov}

With increasing pressure, the vibrational picture breaks down, due to increased intermolecular interactions. Crystal eigenvectors do not project on a product of single molecular
eigenvectors any more, but on a product of a linear combination of molecular eigenvectors:
\begin{eqnarray}
\label{eq:1}
  |\Psi_m \rangle & = & \frac{1}{\sqrt{2}}| \tilde{\Psi}^a_m \rangle \otimes | \tilde{\Psi}^b_m \rangle \\
\nonumber
                  & = & \frac{1}{\sqrt{2}}\left[\sum_k c^a_{km} |\psi^a_k \rangle \right] \otimes \left[ \sum_k c^b_{km} | \psi^b_k \rangle \right].                  
\end{eqnarray}                  

In this formula, the molecular eigenvectors $\psi^{\alpha}_k$ that form the basis set on each molecule $\alpha=a,b$  
are equal up to a rotation / translation: $| \psi^a_k \rangle = \mathbf{U} |\psi^b_k \rangle$.
The molecular limit corresponds to the choice: $c^\alpha_{km} = 0~\forall k \neq n$ in Eq.~\eqref{eq:1}.

\begin{figure}
  \centering
  \includegraphics[width=8cm]{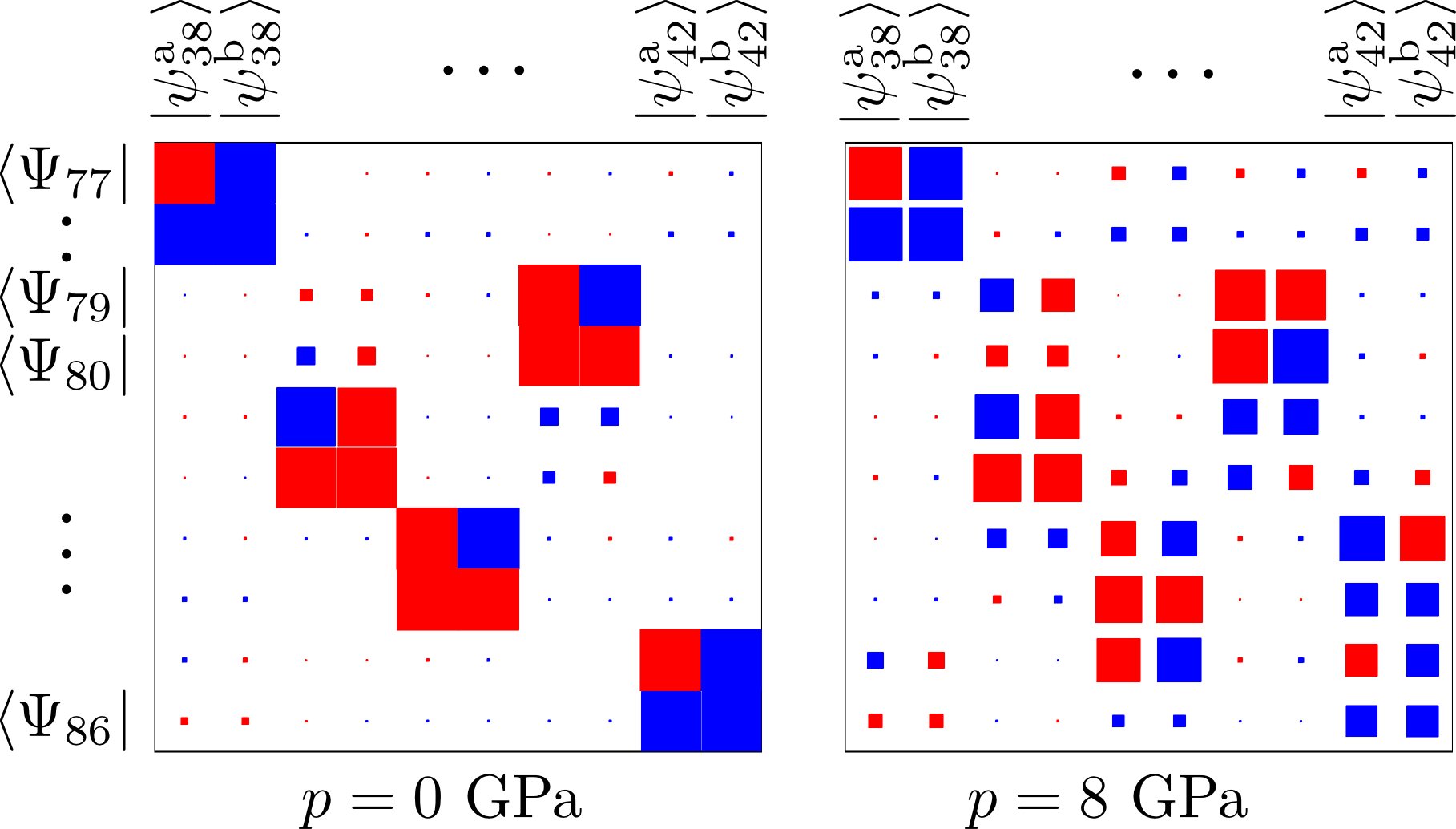}
  \caption{(Color online) Hinton plot for the matrix elements $\Pi_{m n}^{\alpha}$
           of crystal phonon eigenstates $| \Psi_m \rangle$ on
           molecular vibrational eigenstates $| \psi^\alpha_n \rangle$. The area of the coloured square 
is proportional to
           the square of the absolute value of the respective scalar product 
           (see text for details); Red and blue indicate a positive or negative sign of the scalar product ($+/-$). }
\label{figs:9}
\end{figure}

The loss of molecular character with pressure is illustrated in Fig.~\ref{figs:9} using a {\bf Hinton diagram}.
The rows and columns of the matrix represent crystal and molecular eigenvectors -- $\langle \Psi_m |$  and $| \psi^{\alpha}_n \rangle$. 
The size of the squares is proportional to the square of the scalar product between them: $\Pi_{m n}^{\alpha} = \mathrm{sign} (\langle \Psi^\alpha_m | \psi^{\alpha}_n \rangle) \left| \langle \Psi^\alpha_m | \psi^{\alpha}_n \rangle \right|^2$ and $\Pi \equiv |\Pi_{mn}^{\alpha}|$. Red and blue indicate a positive or negative sign of the scalar product, respectively. 
At zero pressure, all the crystal modes in the figure can be associated to a single molecular vibration, in ($+$) or out-of-phase ($-$). 
For example, the crystal mode $| \Psi_{79} \rangle$ is the vibration of two molecular vibrations, out-of-phase, while mode 
$| \Psi_{80} \rangle$ is the superposition of the same two eigenmodes, in-phase. The contribution of the other molecular eigenmodes is negligible. 
At $p=8$~GPa, the situation is not so clear-cut any more. $| \Psi_{79} \rangle$ is now a superposition of two molecular vibrations, with almost equal weight.
This reflects a gradual loss of molecular character with pressure. 

The average molecular projection $\overline{A_{\mathrm{\Pi}}}$ that we used to quantify intermolecular interactions in Fig.~\ref{figs:6}, are defined in terms of these molecular projections $\Pi_{m n}^{\alpha}$:
\begin{equation*}
  \overline{A_{\mathrm{\Pi}}} = \frac{1}{2\cdot N_m} \sum_m \left[\mathrm{max}_n \left(|\Pi^a_{mn}|\right) + \mathrm{max}_n \left(|\Pi^b_{mn}|\right) \right]
\end{equation*}

The residual C$_2$ symmetry enforces $\left| \Pi^a_{mn}\right| = \left|\Pi^b_{mn}\right|$. Because we normalized all \textit{molecular} vibrational states to 1, we get $\left| \langle \Psi_m |  \Psi_m \rangle \right| = 2$. 
This leads to the additional factor 1/2 in front of $\overline{A_{\mathrm{\Pi}}}$.

Being able to define the crystalline eigenvectors in terms of a finite number of molecular modes allows us to follow the evolution of specific modes under pressure. This information is used to define Gr\"uneisen parameters and Davydov splittings.
 Since the molecular character decreases with pressure, this assignment depends very much on the size of the pressure interval between two DFT calculations, as well as on the threshold we choose to define two vectors similar. 

In the following,
 two crystalline eigenstates $| \Psi_j \rangle$, $| \Psi_k \rangle$ are considered as a {\bf Davydov pair} (in and out-of-phase vibrations) if the absolute value of their projection onto the molecular eigenstates is similar (beyond a threshold $T$), and the sign of the projection over one of the two molecules is opposite for the two modes: 
\begin{eqnarray}
\nonumber
  \left|\langle \tilde{\Psi}^a_j | \tilde{\Psi}^a_k \rangle \right| & > & T \\
 -\mathrm{sign} \left(\langle \tilde{\Psi}^a_j | \tilde{\Psi}^a_k \rangle \right) \cdot 
                                      \langle \tilde{\Psi}^b_j | \tilde{\Psi}^b_k \rangle & >&  T
\label{eq:davydov}
\end{eqnarray} 

\begin{figure}
  \centering
  \includegraphics[width=8.5cm]{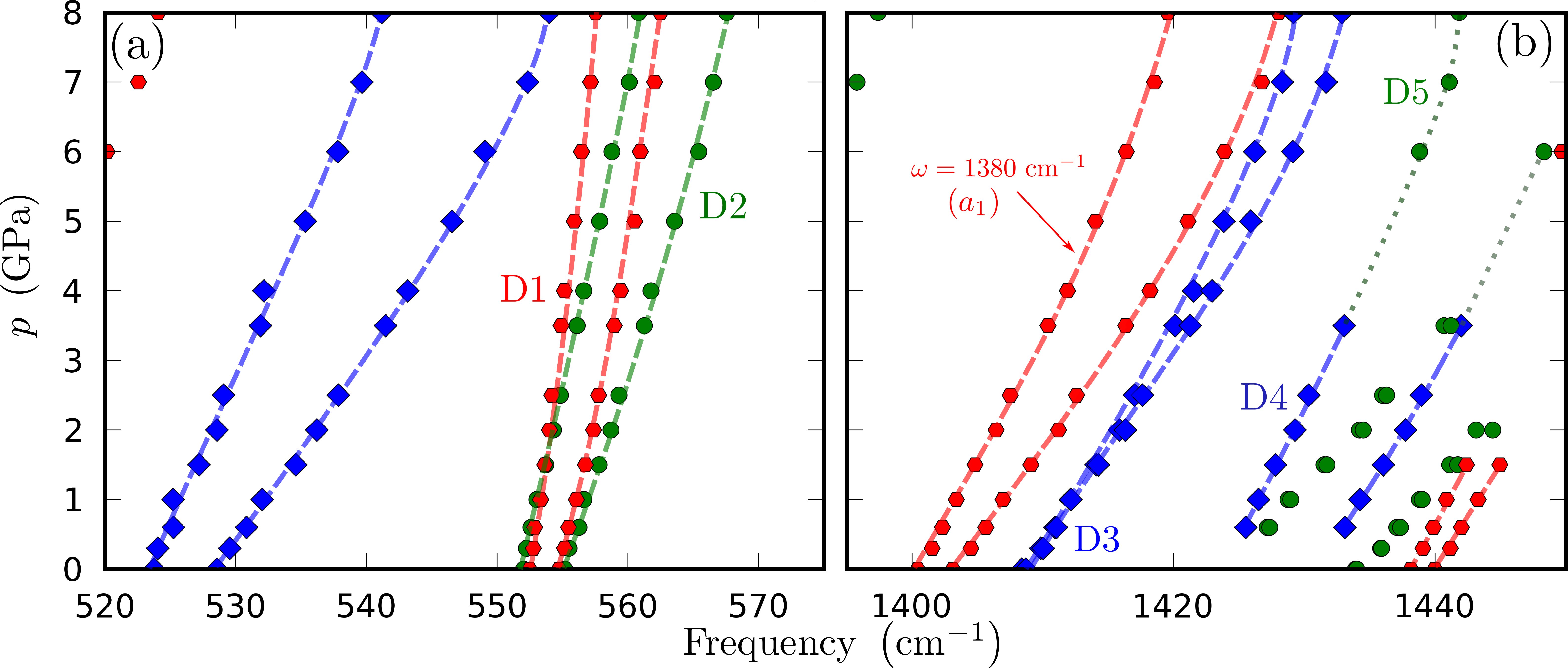}
  \caption{(Color online) Pressure evolution of the frequencies of modes forming a Davydov pair, according to eq.~\ref{eq:davydov}. 
  for two different energy ranges:   (a) $520~\mathrm{cm}^{-1}$ to $575~\mathrm{cm}^{-1}$ and from (b) $1390~\mathrm{cm}^{-1}$ to $1450~\mathrm{cm}^{-1}$.
  Davydov pairs are plotted with the same color/symbol (blue/diamond, red/hexagon, green/circle). The symbols are calculated points and the dashed lines are a guidelines to the eyes.  
 \label{figs:10}}
\end{figure}

In the calculation, we fixed a threshold $T=0.8$ in eq.~(\ref{eq:davydov})  and a threshold of 0.9 for the similarity between different pressures. Overall, we found approximately 
that 160/216 modes at zero pressure form a Davydov pair, indicating that the crystal has a substantial molecular character. 
In Fig.~\ref{figs:10} we plot two different excerpts 
of the whole phonon spectrum over pressure. At zero pressure, the Davydov pair can be either degenerate (e.g. D3) or already split (e.g. D1). With pressure,
the degenerate states split, and their energy differences increases. Some of the frequencies cross ({\em e.g.} D1 \& D2), and some even change 
severely their character -- in Fig. ~\ref{figs:10} ($b$) the pair D5 looks like the continuation for $p \ge 3$~GPa of the pair labelled with D4. 

\section{Conclusions}
\label{sect:conclusion}
In conclusion, in this work we report a very detailed analysis of the vibrational spectra of crystalline picene (C$_{22}$H$_{14}$) under pressure, combining high-quality Raman and IR measurements and state-of-the-art first-principles calculations. In the pressure range we studied (0-8~GPa), DFT calculations reproduce remarkably well the experimental intensities and peak positions. This allowed us to disentangle even the finest details of the experimental spectra, in a system which has 216 phonon modes.
This excellent agrement makes us confident to exploit also the structural results of DFT calculations. These show that in solid picene under pressure the two molecules in the monoclinic unit cell rotate, to reduce the intermolecular distances. Bending of the molecules and intramolecular compression are much smaller effects.

At zero pressure, most phonons of solid picene have a predominant {\em molecular} character; with pressure, due to the increasing intermolecular interaction, many of them acquire a more {\em crystalline} character,  {\em i.e.} they project over more than one molecular eigenmode. This causes an increased complexity of the measured spectra, since more modes acquire a finite IR or Raman cross-section, due to the relaxation of selection rules. 
This evolution is smooth and, even at the highest pressure, picene is still far from the crystalline limit. Following the projections on molecular eigenmodes, it is therefore possible to trace the pressure evolution of individual phonons in the experimental spectra, and to resolve effects which would be otherwise 
impossible to observe, such as the Davydov splitting of the lines.
We found that all phonons display a smooth hardening under pressure,  with Gr\"uneisen parameters of the order of 0.1.

The level of detail of our analysis is to our knowledge unprecedented so far, and can help to resolve several controversies on phonon spectra, electron-phonon coupling and structural effects in K doped picene.~\cite{Casula2011_ele-ph,Subedi2011_ele-ph,Kato2011_ele-ph,Sato2012_ele-ph,Mitsuhash2010,Mitamura2011_perpestive-picene,Kambe_2013}

For example,
on the basis of the calculated Gr\"uneisen parameters, we exclude that K doping leads to a compression of the unit cell, as reported by Ref.~[\onlinecite{Mitamura2011_perpestive-picene}], but rather to an expansion~\cite{Kambe_2013}.
The Gr\"uneisen parameter of the $a_1$ Raman line at 1380~cm$^{-1}$, used as a spectroscopic marker for electronic transfer in K doped picene,
is $\gamma_p$=0.1 for pressure and  $\gamma_d$=0.8 for $x=3$ K doping.~\cite{Kambe_2013}.
This large difference is a clear signature that doping and pressure have distinct effects on the frequency of this vibrational mode. Therefore, in K doped samples, the observed frequency softening cannot be ascribed just to structural effects, but most likely to a strong coupling to electronic degrees of freedom as well.

\appendix
\section{Methods}
\subsection*{Experimental methods} 
Solid picene was prepared by a new optimized synthesis route which permits us to obtain bulk quantities of pure polycrystalline picene powder.~\cite{Malavasi2011_picene-prep} Samples have been fully characterized by X-ray diffraction and ambient pressure Raman and IR spectroscopy in a previous paper.~\cite{JPCM2012_IR_Ram_picene} 
A screw clamped opposing-plate diamond-anvil cell (DAC) equipped with 700 $\mu$m culet II-A diamonds was used to pressurize the samples. The gaskets were made of a 250 $\mu$m thick molibdenum foil with a sample chamber of $\sim$150 $\mu$m diameter and 40-60 $\mu$m height under working conditions. For the IR measurements CsI was used as pressure transmitting medium. High pressure Raman measurements were carried out with and without pressure transmitting medium (NaCl) obtaining actually identical results. During both Raman and IR measurements, pressure was measured in-situ exploiting the standard ruby fluorescence technique.~\cite{Mao1978_Ruby} Further experimental details are reported in Ref.~ \mbox{[\citenum{Marini_exp_2012}]}.

High-pressure IR transmittance data of the picene samples in the DAC were collected at room-temperature exploiting the high brilliance of SISSI beamline of the ELETTRA synchrotron (Trieste, Italy).~\cite{Lupi2007_SISSI}
The incident (transmitted) radiation was focused (collected) by a cassegrain-based Hyperion 2000 IR microscope equipped with a mercury cadmium telluride (MCT) detector for the mid IR range and a He bolometer for the far IR range, both coupled to a Bruker Vertex 70v interferometer which allows us to explore the 400-6000~cm$^{-1}$ spectral range. For the present study, we concentrate on the IR-vibrational modes in the 400-1700~cm$^{-1}$ spectral range.
A thin sample slab obtained by pressing finely milled sample powder between the diamond anvils has been placed on top of a presintered CsI pellet in the DAC gasket hole, thus allowing a clean sample-diamond interface.~\cite{Sacchetti2006_DAC} The slits of the microscope were carefully adjusted to collect only transmitted light from the picene pellet. Their configuration was fixed for all the experiment. 

Two different instruments with different excitation lines have been used for Raman measurements owing to the presence of an unexpected strongly pressure dependent fluorescence emission. As a matter of fact, using the $\lambda = 632.8$~nm excitation line the fluorescence signal prevented the collection of data above $ p \approx$~3~GPa. For this reason we repeated the Raman measurements using a different apparatus with a $\lambda = 785$ nm laser which allowed to measure a good Raman signal up to about 6~GPa. Preliminary experimental studies show that the large fluorescence signal of picene can be ascribed to a two photon process \cite{Citroni2013, Fanetti2013_HP} and that its dependence upon volume compression can be associated to a pressure induced narrowing of the band gap. This is an important point and further experimental and theoretical investigations are in progress. 

The two different confocal-microscope Raman spectrometers were a LabRam Infinity by Jobin Yvon at the Department of Physics of the Sapienza University of Rome and a Raman Senterra by Bruker Optics at Porto Conte Ricerche laboratory (Alghero, Italy). The former was equipped with a He-Ne laser ($\lambda=632.8$~nm) and a 1800~lines-mm$^{-1}$ grating monochromator whereas the latter was equipped with a diode laser ($\lambda=785$~nm) and a 1200~lines-mm$^{-1}$ grating monochromator. In both cases Raman measurements were carried out on powder samples in the backscattering geometry with a spectral resolution better than 3~cm$^{-1}$. A notch filter was used to reject the elastic contribution and a charge-coupled-device (CCD) to collect the spectrum of the scattered radiaton. Using a few micron-sized laser spot on the sample, accurate Raman spectra were collected.

\subsection*{Computational methods}
We calculated the Raman and IR spectra of picene in its crystalline form as a function of pressure using Density Functional Perturbation Theory (DFPT)~\cite{Baroni2001_DFPT,Lazzeri2003_Raman_DFT}, as implemented in the {\em quantum-espresso} package\cite{qe} employing LDA norm-conserving pseudopotentials~\cite{psps}. 
The plane-wave cutoff energy was set to 100~Ry for all calculations.
For the self-consistent calculations, a uniform $2 \times 2 \times 2$ grid was used for $\mathbf{k}$-space integration.
We checked the choice of the cutoff by performing calculations under pressure keeping the number of plane waves constant by increasing/decreasing the cutoff according to $E_{\mathrm{cutoff}} \propto V^{-\frac{2}{3}}$.

For selected pressures, we calculated the Raman and IR spectra. These were then decomposed through a detailed symmetry analysis that employs projections of the phonon eigenvectors ( $\Psi_i$) on the molecular ones ($\psi^{\alpha}_j$). As the lattice structure and the atomic positions are changing with applied pressure, we mapped the two C$_{22}$H$_{14}$ molecules of the crystal unit cell separately onto the molecular calculation using rotations and translations. 

\section{Analysis and discussion of selected spectral ranges}
To illustrate the line-shape analysis of our data, we have selected two spectral ranges (one for Raman and one for IR) where, exploiting theoretical information, it is possible to perform a deeper and accurate analysis of the spectral structures evolving under pressure.  
Note that we already presented in the main text the detailed analysis performed in two cases, namely the 1380~cm$^{-1}$ $a_1$ Raman peak and the IR
400-480~cm$^{-1}$ spectral range.

\subsection*{Raman: 570-790~cm$^{-1}$ spectral range}
\begin{figure}[t]
\includegraphics[width=7.5cm]{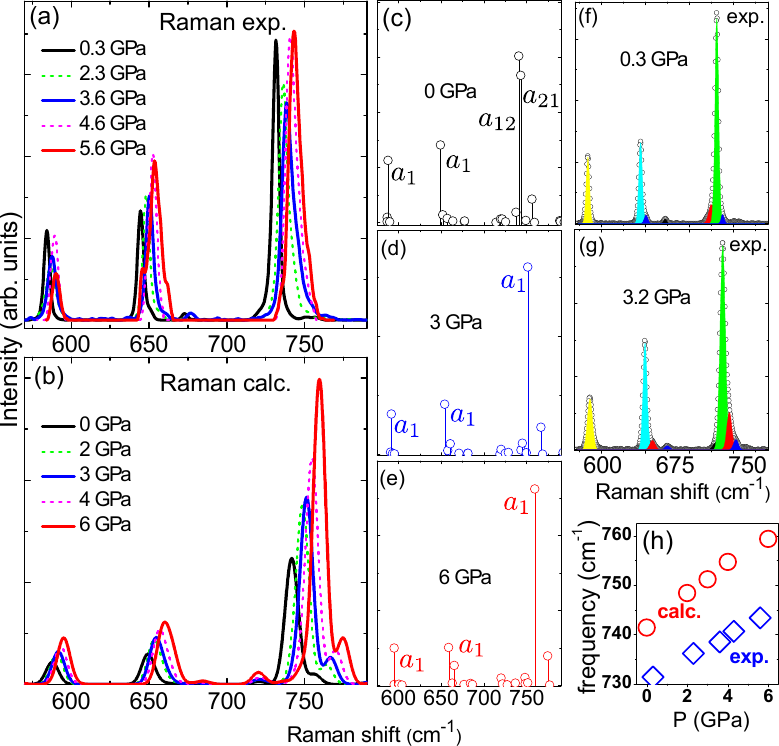}
\caption{(Color online) Raman modes of crystalline picene around 650~cm$^{-1}$ at different external pressures. The experimental data are presented in panel (a). The computed spectra are presented in panel (b) are obtained by convolving the DFT cross sections (see panels (c)-(e)) with a Lorentzian function to mimic 
the experimental broadening. Examples of the spectral deconvolution at two pressures are shown in panels (f) and (g). 
Panel (h) shows the pressure dependence of the frequency of the component which has the highest intensity in experiment (blue)
and theory (red). For the major calculated peaks, the symmetry of the main molecular modes are given (see panels (c)-(e)). Linear combinations 
are shortened as e.g. $a_{121}b_{12} = a_1 + a_2 + a_1 + b_1 + b_2$ whereat the order is descending according to their contribution.}
\label{figs:11}
\end{figure}
Experimental and theoretical Raman spectra within the 570-790~cm$^{-1}$ frequency range are shown in Fig.~\ref{figs:11} at selected pressures. The experimental data (panel (a)) and DFT calculations (panel (b)) show three well isolated spectral structures formed by several components. The spectrum in (b) is obtained from the DFT cross-sections (shown at three selected pressures in panels (c)-(e)) with a convolution of a Lorentzian function to mimic the experimental broadening. The experimental/theoretical comparison demonstrates the excellent level of the agreement: DFT not only captures the main features of the experimental spectra, but also reproduces the observed pressure evolution of the spectral shape, including several minor peaks. This can be better understood looking at the spectral lines around 650 and 730~cm$^{-1}$, where only one component can be observed at ambient pressure 
(see panels (c) and (f) of Fig.~\ref{figs:11}). Exploiting the calculated spectra in Fig.~\ref{figs:11}(d) it is possible to assign pressure-induced lineshape modifications (see Fig.~\ref{figs:11}(g)) to low intesity peaks with a strong pressure dependence. These indeed cannot be resolved at low pressure, but on volume compression they clearly emerge as shoulders of the main peaks. 
We notice that the experimental pressure dependence of the central frequency of the most intense peak is in good agreement with the calculated one, apart from a constant offset~of~$\sim 10$~cm$^{-1}$ (see Fig.~\ref{figs:11}(h)).

\subsection*{Infrared: 1390-1500~cm$^{-1}$ spectral range}
\begin{figure}[t]
\vspace{0.7cm}
\includegraphics[width=7.5cm]{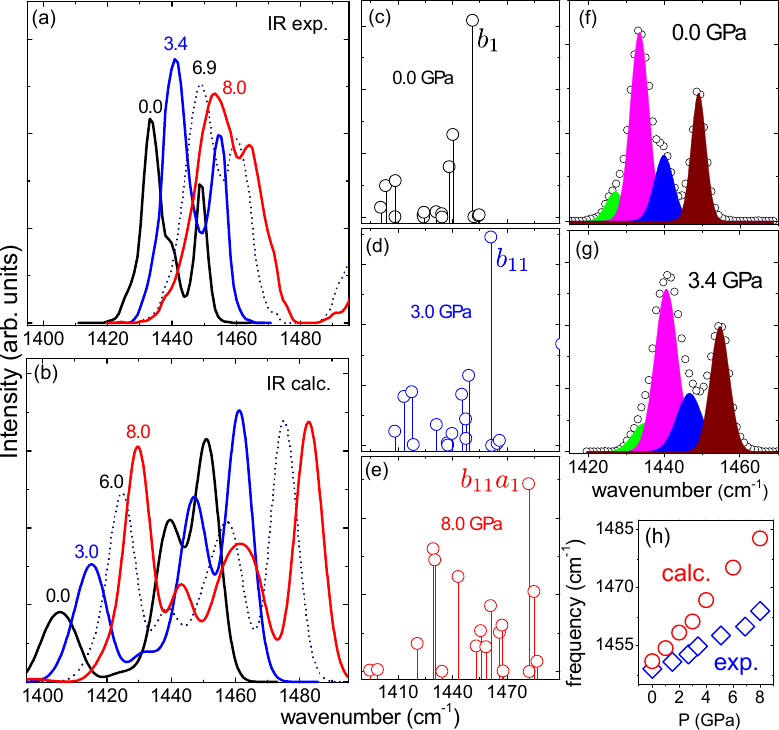}
\caption{(Color online) The IR modes of the picene around 1440~cm$^{-1}$ at different external pressures. Panels are organized according to the same scheme as in Fig.\ref{figs:11}.}
\label{figs:12}
\end{figure}
The pressure evolution of the IR band around 1440~cm$^{-1}$ is shown in Fig. \ref{figs:12}. A huge number of vibrational modes do contribute to the band and a one to one identification between observed and calculated lines is therefore quite difficult. Once again, the reliability of our DFT calculations allowed us to deconvolve the zero pressure spectrum with four main components (see Fig. \ref{figs:12}(c) and (f)). The remarkable broadening observed on 
increasing pressure (see Fig. \ref{figs:12}(g) for data at 3.4 GPa) can be ascribed to usual lattice compressione effects. However, present calculated data indicate that a significant contribution is also due to the pressure driven enhancement of several low intensity modes (compare panels (c), (d), and (e)).
It is worth to notice that, even with the best available spectral resolution, it is impossible to resolve such a large number of almost degenerate modes, as can be seen in Fig. \ref{figs:12}(d). A limit of the analisys is shown in Fig. \ref{figs:12}(h), where a different pressure evolution of the frequency of the most intense calculated mode and the corresponding fourth component in the measured spectrum can be observed. This may indicate that, at variance with the other above mentioned cases, it is difficult to establish a one to one correspondence between calculated and measured peaks in this energy range.


\begin{thebibliography}{99}
\bibitem{note:tab:crit}{The modes with at least 15\% of the peak cross section and their Davydov partner, if it exists, have been taken.}
\end{thebibliography}
\end{document}